\documentclass[10pt,conference]{IEEEtran}

\makeatletter
\newif\if@restonecol
\makeatother

\usepackage{amsthm}
\usepackage{mdframed}
\usepackage{url}
\usepackage{paralist}
\usepackage{listings}
\usepackage{color}
\usepackage{transparent}
\usepackage{ifthen}
\usepackage[ruled,vlined,linesnumbered]{algorithm2e}
\usepackage{algpseudocode}
\usepackage{flushend}
\usepackage{multirow}
\usepackage{xspace}
\usepackage{amsfonts}
\usepackage{amsthm}
\usepackage{subfig}
\usepackage{expdlist}
\usepackage{tabulary}
\usepackage{framed}
\usepackage{epstopdf}
\usepackage{colortbl}
\usepackage{graphicx} 
\usepackage{amsmath}
\usepackage{xifthen}
\usepackage{xcolor}

\usepackage{makecell}

\usepackage{siunitx}
\usepackage{cleveref}

\usepackage{tikz}
\usepackage[nocompress]{cite}
\usepackage{balance}
\usepackage{flushend}

\newmdtheoremenv[skipabove=\baselineskip,skipbelow=\baselineskip]{boxedObservation}{Observation}

\renewenvironment{leftbar}[1][\hsize]
{\MakeFramed{\hsize#1\advance\hsize-\width\FrameRestore}}
{\endMakeFramed}

\definecolor{codegreen}{rgb}{0,0.6,0}
\definecolor{codegray}{rgb}{0.5,0.5,0.5}
\definecolor{codepurple}{rgb}{0.58,0,0.82}
\definecolor{codebackcolor}{rgb}{0.95,0.95,0.92}

\algnewcommand{\AND}{\textbf{AND}\xspace}
\algnewcommand{\OR}{\textbf{OR}\xspace}

\newcommand{\ie}{\textit{i.e.,}\xspace}
\newcommand{\eg}{\textit{e.g.,}\xspace}
\newcommand{\etal}{\textit{et al.}\xspace}

\def\validator{\textsc{Commons-Validator}\xspace}
\def\money{\textsc{Joda-Money}\xspace}

\def\spotify{\textsc{Spotify-Web-Api}\xspace}
\def\cdk{\textsc{Cdk-Data}\xspace}

\def\cli{\textsc{Commons-CLI}\xspace}

\def\jfreechart{\textsc{JFreeChart}\xspace}

\def\reader{\textsc{Jline-Reader}\xspace}

\def\RQa{\emph{RQ1}\xspace}
\def\RQb{\emph{RQ2}\xspace}
\def\RQc{\emph{RQ3}\xspace}
\def\RQd{\emph{RQ4}\xspace}

\def\ripr{\textsc{RIPR}\xspace}

\def\analysis{memory-state analysis\xspace}

\newcommand{\mysubsection}[1]{\vspace{1.25mm}\noindent\textit{\textbf{#1.}}}
\newcommand{\myRQsubsection}[1]{\vspace{1.25mm}\noindent\textit{\textbf{#1}}}
\newcommand{\mysubsubsection}[1]{\vspace{1.25mm}\noindent\textit{#1.}}

\newcommand{\CASE}[1]{\STATE \textbf{case} #1\textbf{:} \begin{ALC@g}}
\newcommand{\ENDCASE}{\end{ALC@g}}

\newcommand{\DEFAULT}{\STATE \textbf{default:} \begin{ALC@g}}
\newcommand{\ENDDEFAULT}{\end{ALC@g}}
\newcommand{\DEFAULTLINE}[1]{\STATE \textbf{default:} }

\def\d3js{\textsc{D3.js}\xspace}

 \usepackage[switch]{lineno}

\begin{document}

\title{Leveraging Propagated Infection to Crossfire Mutants}

\author{
    \IEEEauthorblockN{Hang Du}
    \IEEEauthorblockA{University of California, Irvine\\ 
    Irvine, CA, USA \\
    hdu5@uci.edu}
        \and
    \IEEEauthorblockN{Vijay Krishna Palepu}
    \IEEEauthorblockA{Microsoft, Silicon Valley Campus \\ 
    Mountain View, CA, USA\\
    vijay.palepu@microsoft.com}
        \and
    \IEEEauthorblockN{James A. Jones}
    \IEEEauthorblockA{University of California, Irvine\\ 
    Irvine, CA, USA \\
    jajones@uci.edu}
}

\maketitle

\begin{abstract}
Mutation testing was proposed to identify weaknesses in test suites by repeatedly generating artificially faulty versions of the software (\ie \emph{mutants}) and determining if the test suite is sufficient to detect them (\ie \emph{kill} them).
When the tests are insufficient, each surviving mutant provides an opportunity to improve the test suite.
We conducted a study and found that many such surviving mutants (up to 84\% for the subjects of our study) are detectable by simply augmenting existing tests with additional assertions, or \emph{assertion amplification}. 
Moreover, we find that many of these mutants are detectable by multiple existing tests, giving developers options for how to detect them. 
To help with these challenges, we created a technique that performs memory-state analysis to identify candidate assertions that developers can use to detect the surviving mutants.
Additionally, we build upon prior research that identifies ``crossfiring'' opportunities --- tests that coincidentally kill multiple mutants.
To this end, we developed a theoretical model that describes the varying granularities that crossfiring can occur in the existing test suite, which provide opportunities and options for how to kill surviving mutants.
We operationalize this model to an accompanying technique that optimizes the assertion amplification of the existing tests to crossfire multiple mutants with fewer added assertions, optionally concentrated within fewer tests.
Our experiments show that we can kill \emph{all} surviving mutants that are detectable with existing test data with only 1.1\% of the identified assertion candidates, and increasing by a factor of 6x, on average, the number of killed mutants from amplified tests, over tests that do not crossfire.
\end{abstract}

\section{Introduction}
\label{sec:introduction}

The ultimate goal of mutation testing is to allow software developers to create stronger test suites.
It does this by injecting artificial faults (\ie mutations) into a program to identify weaknesses in the test suite (\ie mutations that are not detected).
A developer would then write tests to detect (or \emph{kill}) the undetected or \emph{surviving} mutants.
The intuition of this approach is that by strengthening the test suite to detect the mutations, the test suite will then be more likely to catch future \emph{real} faults before they can cause any adverse effects upon users. 

Multiple works on test generation and amplification (\eg \cite{fraser2010mutation,Fraser2013WholeTestSuiteGeneration,Bandry2015Dspot,danglot2019dspot,Souza2014TestDataMutationTesting,Barr2015OracleProblem,Danglot2019Amplification}) focus on automating the improvement of test suites. Some of these works address generating specific parts of the test suite, such as test data and test oracles (\eg \cite{Souza2014TestDataMutationTesting,Anand2013SurveyTestDataGeneration,Barr2015OracleProblem}), while others generate entire test suites from scratch (\eg \cite{Fraser2013WholeTestSuiteGeneration,Fraser2011EvoSuite}). 
Additionally, some techniques \emph{amplify} existing test suites by exploring new test data and assertions (\eg \cite{danglot2019dspot,Danglot2019Amplification}). These approaches enhance test suites and are evaluated based on their ability to produce higher mutant-killing ratios.
However, the usage scenarios of these techniques often differ from typical mutation-testing practices in either or both of the following two key ways: (1) Mutation testing practitioners target individual surviving mutants and incrementally improve existing test suites \cite{petrovic2018state,Petrovic2023FixThisMutant,beller2021would,smith2009guiding}, and (2) they perform mutant-killing activities on pre-constructed, human-written test suites that already contain test data and oracles \cite{beller2021would,Petrovic2023FixThisMutant,petrovic2018state,Rojas2017GamifyCroudsource}.

From (1), we recognize the importance of analyzing specific surviving mutants to help practitioners target and kill individual mutants. From (2), we acknowledge the presence of multiple existing developer-written tests that may execute the mutant, creating opportunities for test amplification---if a slight improvement to an existing test can kill a surviving mutant, there is no need to design a different test from scratch.

The classic fault-error propagation model, RIPR model \cite{Li2017TOS,Li2014OracleConference}, investigates such scenarios where a test case executes a specific fault. The model includes four conditions: the fault must be executed (\emph{R}eachability), infect the program states (\emph{I}nfection), propagate the infection (\emph{P}ropagation), and have appropriate test oracles to reveal the fault (\emph{R}evealability).
The last condition of revealability relies on the test case having appropriate and sufficient test oracles. If it does not, adding additional assertions, or \emph{assertion amplification} may help reveal the fault. 
Following the \ripr model, Du~\etal\cite{Du2024Ripples} empirically investigates the end-to-end runtime effects of mutation execution, which uncovers opportunities to kill surviving mutants through such amplification.

Furthermore, while targeting a specific surviving mutant, mutation-testing researchers discovered that a human-designed test for one surviving mutant sometimes coincidentally kills other surviving mutants---a phenomenon termed ``crossfire'' \cite{SmithWilliam2007Crossfire,smith2009guiding}.
Understanding and leveraging these mechanisms can strengthen each incremental test-augmentation (mutant-killing) attempt.

In this work, we (1) offer a model to investigate the causes and intricacies behind the crossfire phenomenon in both the mutation-analysis and mutation-testing processes, (2) empirically analyze existing test suites' mutant-crossfire capabilities at both the test and assertion granularities, (3) systematically investigate assertion-amplification opportunities for each surviving mutant, (4) develop techniques that recommend mutant-crossfiring assertions with crossfiring goals at varied granularities, (5) compare and evaluate our assertion-amplification techniques of varied crossfiring strategies, and (6) gain initial insights into mutant killing assertion candidates' characteristics.

Through our analysis, we found varied mutant-killing capabilities of individual assertions and test cases across different projects, unveiled how passing test runs exhibit propagation, and discovered overwhelming surviving mutant-killing opportunities through assertion amplification.
Our surviving mutant-crossfiring techniques allow for a small selective set of assertion-amplified, developer-written tests to crossfire a substantially larger number of surviving mutants, by a factor of $6.1$.

The main contributions of this paper include:

\begin{itemize}
	\item An analysis technique that assesses the granular infected memory locations that resulted from the propagation of the infection caused by execution of the injected faults. This technique offers recommendations for additional assertions that can be used to kill surviving mutants.
	\item A theoretical model that dissects and analyzes propagation at fine-grained levels and illustrates the ``crossfire'' effects in mutant kills and a technique that provides optimizations for adding more effective assertions to tests.
	\item An empirical evaluation of our proposed techniques showing the varied capabilities of tests and assertions for detecting propagated state infection, assessing our ability to recommend assertions to kill surviving mutants, and assessing our optimizations to crossfire mutants.
	\item An implementation and dataset to allow for future research and experimental reproducibility.
\end{itemize}

\section{Motivation and Challenges}
\label{sec:motivation}

\begin{figure*}[tbh]
\centering
\includegraphics[width=\linewidth]{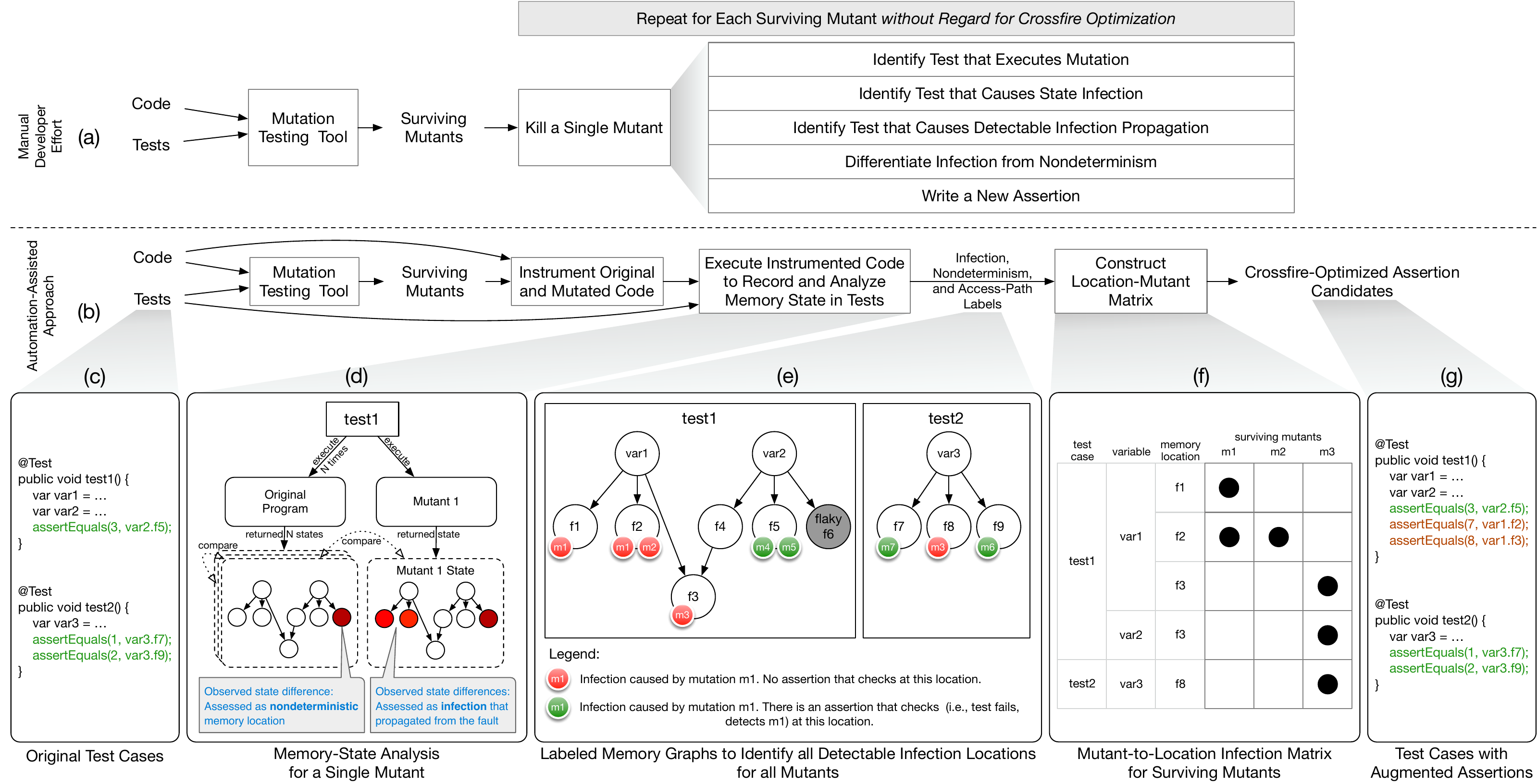}
\caption{Process to kill surviving mutants by manual developer effort (top) that does not optimize for crossfiring due to its piecemeal approach, and an automation-assisted approach (bottom) that analyzes state infection and optimizes for crossfiring.\label{fig:graph}}
\end{figure*}

Mutation testing involves both test automation and developer effort. 
A tool like \textsc{PIT}~\cite{PIT} or \textsc{$\mu$Java}~\cite{ma2005mujava} typically handles the automated portion. 
These tools inject faults, rerun the test suite for each mutant, calculate the mutation score, and report unkilled mutants. 
The output of the tool provides developers with two key benefits: (1) an assessment of the test suite's strength through the \emph{mutation score}, and (2) a prescription on improving the test suite via a list of surviving mutants. 
With these results, developers must then strengthen the test suite by targeting surviving mutants. 
The manual process that a developer might take to kill surviving mutants is depicted in the top row of \Cref{fig:graph}.

Once a mutation-testing tool is used to reveal surviving mutants, a developer may wish to kill one surviving mutant by first identifying a test that executes the mutation. 
The RIPR model~\cite{Li2014OracleConference,Li2017TOS} prescribes the execution (or \emph{Reachability}) as the first condition to detect a fault.
Next, a developer would want to find a test that not only executes the mutation, but also does so with test data that causes the mutation to affect, or \emph{Infect}, the state. 
Then, a developer would need to create a test assertion to detect an infected state that \emph{Propagated} back to the test, in order to \emph{Reveal} the mutation or fault and kill the mutant.

Each of these steps provides challenges for a developer. 
To find such infections, a developer might probe the state (through print statements or a debugger) to determine if any part of the state that is accessible from the test case shows any infection, as evidenced by differences with the original (unmutated) program's state from the same test.
Even once such differences in state are found, another challenge is determining if those differences are caused by the infection from the mutation or from nondeterministic values (such as a date or time-of-day field, a hashcode, or a thread identifier).
Even once a real infection has been identified, there may be multiple ways to access that infection, and the developer would need to make a choice as to how to access it in order to detect it with an assertion. 
Moreover, this entire process must be repeated for each surviving mutant. 
Mutation-testing tools often report hundreds or thousands of surviving mutants, even for mature and well-tested software \cite{petrovic2018state}. 

This last challenge of the magnitude of the problem could be somewhat alleviated by the phenomenon of ``crossfiring'' mutants~\cite{SmithWilliam2007Crossfire,smith2009guiding}---a test for one surviving mutant sometimes coincidentally kills other surviving mutants.
However, such a manual, mutant-by-mutant approach would likely not exploit the crossfire effects to their true potential. 

\vspace{1.25mm}
Based on the potential opportunities and these observed challenges for developers to perform the manual portion of analyzing each surviving mutant, our motivation is to provide conceptual models and analyses to help researchers understand the various aspects of this task, as well as to create practical techniques to help developers kill surviving mutants with greater assistance. \section{A-model: A Breakdown of Mutant Crossfire Effects}
\label{sec:model}
\begin{figure}[tb]
\includegraphics[width=0.5\textwidth]{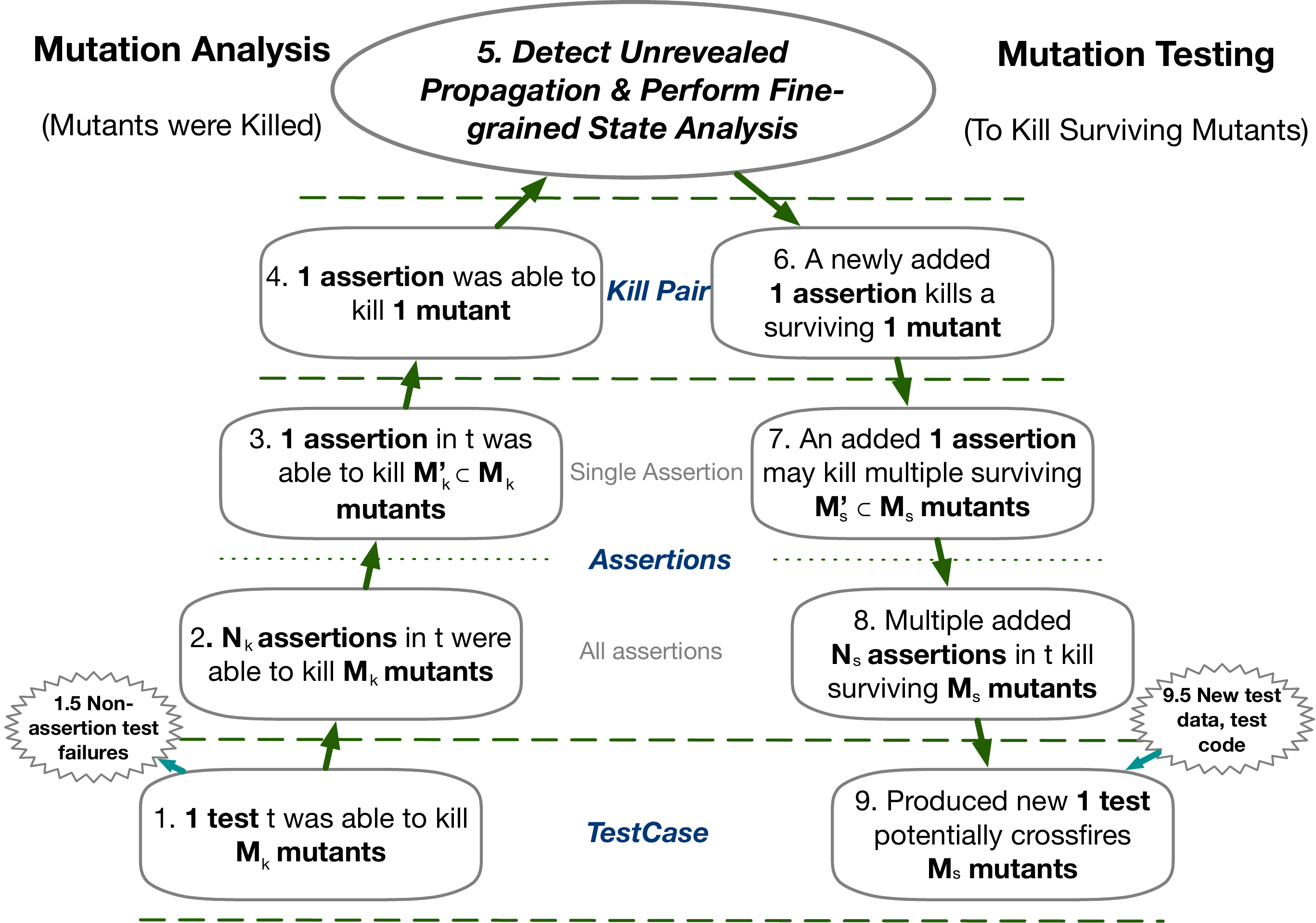}
\caption{A-model: A Breakdown of Mutant Crossfire Effects}
\label{fig:A}
\end{figure}

To understand the natural occurrence of the mutant ``crossfire'' phenomena, and to exploit its potential in killing surviving mutants, we developed a model in \Cref{fig:A} to illustrate the observed mutant crossfire effects, as a result of intricate fine-grained memory infection, in the mutation-analysis phase and the mutation-testing phase.

The model's structure takes the shape of the letter ``A,'' which symbolizes the divergence of two mutation phases: an existing test case is evaluated on killed/detected mutants during the mutation-analysis phase on the left leg, and subsequently, this test case may be augmented with assertions in the mutation-testing phase to kill surviving mutants on the right leg.
We start by explaining the A-model's left leg in the mutation-analysis phase by breaking down an existing test, using the example shown in the bottom row of \Cref{fig:graph}.
The example includes both the original test (\Cref{fig:graph}(c)) and the augmented test (\Cref{fig:graph}(g)), and an anatomy of mutation effects in the intermediate subfigures (\Cref{fig:graph}(d)-(f)).

In Block 1, an existing test case was executed on mutants and demonstrates its capability to kill mutants. 
For instance, \textit{Test 1} in \Cref{fig:graph}(c) killed two mutants (\textit{m4} and \textit{m5}).
When a test kills mutants, it can fail by assertions (Block 2) or non-assertion failures (such as crashes \cite{ToKillAMutant2023Hang}) (Block 1.5).
\textit{Test 2}'s set of assertions cumulatively killed two mutants (\textit{m6} and \textit{m7}).
As a test case may contain multiple assertions, each assertion may be capable of killing multiple mutants (Block 3). 
\textit{Test 1}'s assertion of \texttt{var2.f5} detects two mutations (\textit{m4} and \textit{m5}).
As a result, an individual assertion contributes to a mutant kill (Block 4), creating a set of assertion-mutant pairs (\eg \texttt{var2.f5}-m4).

At the apex of the model, an analysis can reveal propagated infection despite the fact that the tests pass (Block 5).
As described in \Cref{sec:motivation}, performing such analysis can be difficult and time consuming.
However, an automated analysis may be able to identify infections that propagate from a mutation back to a test, and as such suggest an assertion candidate to kill it (Block 6).
In \Cref{fig:graph}(e), for instance, if we add an assertion in \textit{Test 1} that asserts the memory location at node \textit{f3}, then the surviving mutant \textit{m3} will be killed.
Moreover, each added assertion may kill multiple surviving mutants (Block 7).
For instance, in \Cref{fig:graph}(e), an assertion for memory location \textit{f2} in \textit{Test 1} crossfires surviving mutants \textit{m1} and \textit{m2}, because those two mutants trigger infection at the same memory location. 
Furthermore, a test may be augmented with multiple assertions (Block 8), and each of them may kill some surviving mutants.
As a result, the newly produced test (\eg \textit{Test1} in \Cref{fig:graph}(g)) with assertion amplification exhibits its capability of killing multiple surviving mutants (Block 9). 
Note that, such mutant kills may also be produced by crafting a brand new test case with new test data and execution logic (Block 9.5).

Throughout the A-model, we demonstrate the crossfire effects of an existing test case from the individual test case and individual assertion level, as a result of assessing granular memory infections for multiple mutants. 
Also, we recognize the potential to kill multiple surviving mutants at the test- and assertion-level in the test-augmentation process.
Moreover, such assertion-amplification opportunities may occur across multiple existing tests for each surviving mutant.
In the next section, we present our technique to strategically augment an existing test suite with new assertions to kill surviving mutants according to this crossfire-aware model, and thus offer developer automated assistance to strengthen their test suite.

 \section{Approach}
\label{sec:approach}
In this section, we describe our technique to produce assertion candidates for strategically killing surviving mutants. Our technique can be decomposed into the following steps:
\begin{enumerate}
	\item Run mutation-testing tool on program $P$ with its test suite $T$ to get the list of surviving mutants $M$.
	\item Instrument $P$ to record all reachable memory states for each test case in $T$. The output is the instrumented program $P'$.
	\item Execute the instrumented program $P'$ $N$ times on its test suite $T$ to produce $N$ copies of memory states $S$. 
	\item Perform a comparison across all $N$ copies of $S$. Those memory locations that are consistent across all $N$ copies of $S$ are labeled as deterministic $S_{P,d}$, and those locations that show any differences are labeled as nondeterministic $S_{P,n}$.
	\item For each surviving mutant $M_i$ in $M$, instrument mutant $M_i$ to record all memory states for every test case. The output is instrumented mutant program $M_i'$.
	\item Execute instrumented mutant $M_i'$ with test suite $T$ to record all state for all test cases $S_{M_i}$.
	\item Filter the state of the mutant $S_{M_i}$ to remove all nondeterministic locations $S_{M_i,d}$.
	\item Compare deterministic state of the original program $S_{P,d}$ with the deterministic state of the mutant $S_{M_i,d}$ to identify all locations $S_{M_i,i}$ that reveal infection.
	\item Build a matrix $\mathbb{M}$ for all memory locations that reveal any infections $S_{M_i,d}$ across all surviving mutants $M$.
	\item Performing strategies on $\mathbb{M}$ to optimize crossfiring to produce a list of assertion candidates, and for each assertion candidate the list of surviving mutants that it would kill.
\end{enumerate}

This process can be summarized into two primary stages: (a) a fine-grained memory-state analysis on surviving mutants, through which we analyze the location of state infection in memory and the ways to access the pollution so as to derive assertion candidates (Steps 2--8), and 
(b) assertion-candidate selection, where we leverage analyzed granular infection behaviors to design assertion-candidate selection strategies that crossfire mutants (Steps 9--10).
This process is depicted in the second row of \Cref{fig:graph} along with example test cases, memory states, memory-location-to-mutant matrix, and resulting augmented test cases.

\subsection{Memory Analysis and Assertion Candidate Generation}
\label{sec:step1}
A new assertion may be derived to check a specific, previously unchecked granular program-state behavior through accessing variables that are directly exposed from the test-case scope.
As such, we analyzed each variable's program states for each passing test run (\Cref{fig:graph}(d)). 
Note, in this work, we refer to ``variables'' as any local variables, fields of test-class instances, method-return values, instantiated objects, or static fields---that is, any memory location that is immediately accessible from each test case's scope. 
For each variable's program-state analysis, we performed $N$ repetitive test runs on the original program to identify nondeterministic memory locations (such as node \texttt{f6} in \Cref{fig:graph}(e)), and compared their states to their counterpart from a mutant, while ignoring non-deterministic locations.
We used breadth-first search in the traversal of each pair of matched memory graphs to perform such a comparison, starting from the root variable.

Each node in the object graph represents an instance of an object (or primitive value at the leaves) and is given a unique node ID.
Each edge indicates a relation between objects: an element from a size-changeable collection (\eg array, list) or a field of an object.
During graph traversal, any granular memory differences detected at a node is marked, and the traversal along that path ceases any further comparisons deeper along that path.
Meanwhile, graph structural differences may occur during comparison, which can result from null fields and mismatches in collection sizes.
If a collection-size infection occurs, further state comparison under the collection is stopped.
Therefore, the algorithm yields a set of granular state-difference details between two object graphs.
Each of them includes a unique node ID that specifies the location of the infection relative to the object graph.

Such an analysis is performed on all surviving mutants' covering test runs and individually on each variable. 
To mitigate the redundant calculation and analysis of variables pointing to the same object in memory, we apply a hashing algorithm on variable-level object graphs.
As a result, this approach yields a set of granular state-difference details between two object graphs.
Each of them includes a unique node ID that specifies the location of the infection retrievable from an object graph.

As a result, we produce comprehensive sets of program-state infection locations for surviving mutants: each of them includes the surviving mutant ID, test case ID, variable ID, and node ID accessible via the variable pinpointing the exact pollution (\Cref{fig:graph}(f)).
This information forms a set of assertion candidates with each specifying which surviving mutant can be killed, which test case can be augmented, which variable in the test case can access the pollution, which part of the object graph should be asserted, and what are their expected and polluted values.

\subsection{Assertion Candidate Selection and Mutant Crossfire}
\label{sec:step2}
To reduce the inevitable human-in-the-loop engineering cost, only a few assertions are required to kill all of the assertion-actionable mutants.
For example, some assertions may be written with shorter access paths than others.
Specifically, we measure the depth of the polluted node accessible from the corresponding variable starting from depth of 1.
In \Cref{fig:graph}(e), node \textit{f3} has a depth of 3 relative to \textit{var 2} and a depth of 2 relative to \texttt{var 1}.
As such, an assertion-checking memory location \textit{f3} may be easier to access from variable \texttt{var 1}, thus potentially making the assertion more readable with less comprehension and engineering cost.
Moreover, we recognize that practitioners often perform incremental augmentation and validation to kill surviving mutants.
Focusing these efforts within a few locations in the test code can reduce the engineering and validation burden, minimizing the need to switch contexts between different amplification locations in the test code.
As such, we can minimize the number of assertions (\eg checks for \textit{f2} and \textit{f8}), minimize the number of variable checks (\eg checks for \texttt{var 1}) and minimize the number of tests to perform such assertion amplification (\eg improve \texttt{Test 1} only rather than both tests) while achieving the same mutation score.

\begin{figure}
\includegraphics[width=0.5\textwidth]{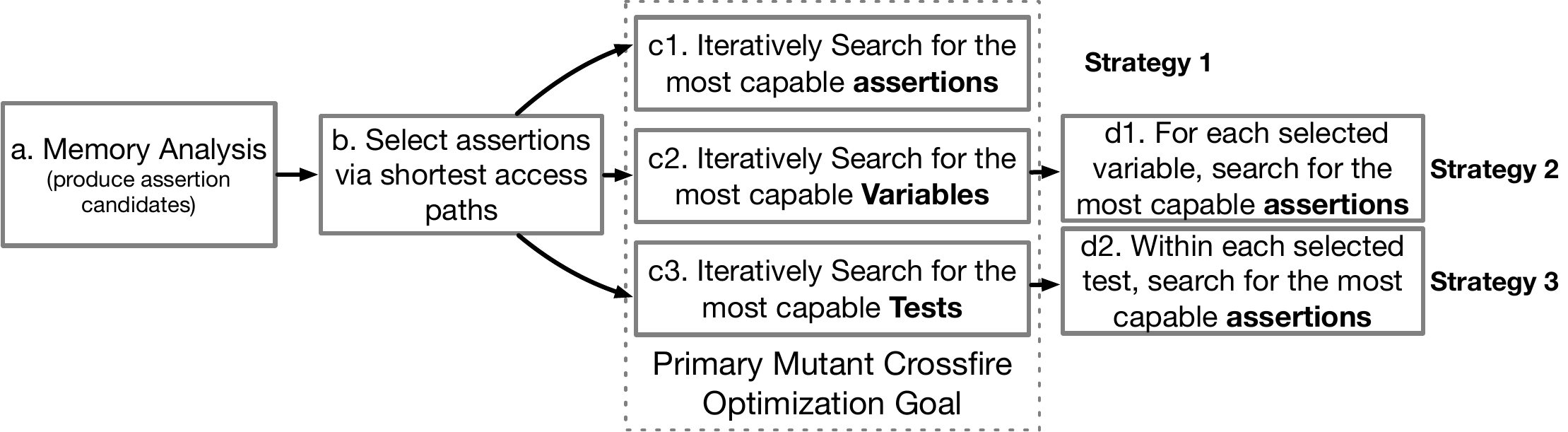}
\caption{Mutant Killing Strategies}
\label{fig:crossfire}
\end{figure}

As such, we presented three different mutant crossfiring strategies shown in \Cref{fig:crossfire}.
In Block b, an initial optimization heuristic can be applied to select candidate assertions by filtering out the assertion candidates via non-shortest access paths for each surviving mutant.
This filtering is based on the heuristic that assertions checking attributes deeply accessible from a variable are more likely to be brittle, which may reduce the maintainability of the test suite.
Then, we apply greedy heuristics---iteratively search for the assertion/variable/test that kills the largest number of surviving mutants (c1, c2, and c3) among all candidate assertions until all assertion-actionable surviving mutants are killed.
Such greedy heuristics prioritize assertions, variable checks, or tests that should be augmented based on their ability to kill the largest number of mutants, thereby limiting the human-engineering burden.
Similarly, greedy heuristics could be applied further on non-finest searching scopes, \ie variables and tests, where we continue to iteratively search for the most capable assertions related to a specific variable (d1) or test (d2), corresponding to Strategy 2 and Strategy 3.

As a result, our technique produces assertion candidates that not only kill but also crossfire surviving mutants, which requires fewer and more concentrated updates in the existing test suite to strengthen the test suite.

\section{Evaluation}

\label{sec:experiment}
In this section, we outline the experimental design to assess the effectiveness of our approach for generating assertion candidates that kill surviving mutants, while optimizing for crossfiring assertions.
We enumerate the research questions that will guide our empirical investigations.
To answer those questions, we implemented our approach, performed experimentation on 10 popular open-source Java programs, and report our findings in \Cref{sec:results}.

\subsection{Research Questions}
\label{sec:rqs}

These research questions are structured to assess both legs, or facets, of the A-model: (1) how test assertions detect infection (\ie kill mutants); and (2) how assertions and tests crossfire multiple mutants.

\myRQsubsection{\RQa: For killed mutants, how do tests and assertions contribute to the test suite' fault-detection capabilities?}
Tests often contain multiple assertions that together reveal infections through test failures.
However, the fault-detection capabilities of individual assertions in a test may vary.
For instance, a single test with two assertions may kill three mutants, but with only one assertion killing all three mutants.
For a developer, this may potentially make some assertions more useful than others.
In a different example, a single test contains three assertions, each of which kills a unique mutant, and as such each demonstrate their utility.
We investigate this phenomenon by measuring: (a) the count of mutants killed by each test, (b) the count of mutants killed by each assertion, and (c) the count of assertions in each test.

\myRQsubsection{\RQb: How many surviving mutants are detectable by existing tests, and thus killable by our approach?}
Tests can only detect infections that propagate to the test code's execution scope.
Often, infections may propagate to a test's execution scope, but may go unrevealed because the test lacked an appropriate assertion to detect the infection.
If a developer can detect faults by augmenting an existing test with an assertion, without writing a whole new test case, it might save engineering cost to create new test inputs and identify surviving mutants that are guaranteed to not be equivalent mutants.
We investigate this phenomenon, and report the number of surviving mutants that could be killed by augmenting existing tests with assertions.

\myRQsubsection{\RQc: How many tests offer the opportunity for assertion amplification?}
We investigate the number of tests that do not reveal propagated infections and offer the chance to add newer assertions to kill surviving mutants.
We suspect that developers would ultimately want to (or perhaps even need to) assess generated assertion candidates.
And so, if we augment a greater number of tests with assertions, then that may increase the manual work for a developer. 
When manually assessing the generated assertion candidates, the developer would need to understand the logic for more test cases.
In our experiments, we report the number of tests wherein our \analysis was able to generate assertion candidates.

\myRQsubsection{\RQd: How effective are the crossfire strategies at reducing the need for additional assertions to kill surviving mutants?}
As explained from the right leg of the A-model, surviving mutants could be killed and even crossfired.
Our technique searches for assertions, test variables, or test cases that guide the killing of a maximal number of mutants to achieve the best crossfire effects.
Developers may need to manually assess generated assertion candidates.
In that event, it might be useful if developers need to examine fewer test cases, test variables and assertions while killing a maximal number of surviving mutants.
We evaluate the magnitude of such crossfire effects and capabilities of three different mutant crossfire strategies. 
We further compare them and discuss the trade-offs of the three strategies.

\subsection{Experimental Setup}

\mysubsection{Mutation analysis with PIT}
To conduct our empirical analysis and technique evaluation, we employed PIT~\cite{PIT}, a mutation-testing tool for Java extensively utilized in research and practice (\eg \cite{kintis2018effective,shi2019mitigating}).
We used the ``DEFAULTS'' group of mutation operators provided by PIT, which contains a set of mutation operators that has been widely adopted in practice. Each mutation operator in this group ensures one-to-one mappings between the bytecode syntax and the resulting mutation, and pre-filters for bytecode-equivalent mutants \cite{Coles_2023}, thereby mitigating equivalent and duplicate mutants.We modified PIT's source code to enable 10 non-mutation test runs, isolate mutant execution in separate JVM instances, and collect final program states for each test run. 
We also configured PIT to enable the execution of all covering test runs for all mutants.

\mysubsection{Memory Object-Graph Instrumentation}
To collect final program memory states, we instrumented test code (with ASM~\cite{bruneton2002asm}) to identify and record a list of test variables and memory state accessible in the test's execution scope.
We collect the memory object graphs for each local variable, static field, and heap location that is accessible from a test method.
We use XStream \cite{xstream} to record memory object graphs, with configurations to exclude states related to threading and logging utilities.
We customized XStream to support comparison of arrays and collections as part of our fine-grained memory state analysis, detailed in \Cref{sec:step1}.
We also employed static-field cleaners to mitigate state, or test data pollution that might occur across two successive test runs.
These implementation details are captured in our artifact.\footnotemark[1]

\mysubsection{Experimental Steps}
We first ran our analysis to attribute test-failure causes for failing test runs.
Specifically, we count the number of mutants that each test case and each assertion kills.

Next, we compare program states between mutated and original runs on individual variables in the maintained variable list on surviving mutants, which produces comprehensive assertion candidates, as introduced in \Cref{sec:step1}.

Finally, we apply the three mutant-crossfire strategies introduced in \Cref{sec:step2}.
The greedy heuristics used in our strategies may produce multiple equally optimal choices at a given step, thus yielding close-to-optimal but non\-deterministic performance.
As such, we ran each of our strategies 20 times to evaluate their average performance.

\mysubsection{Subject Programs}
We ran our experiments on 10 open-source Java projects, chosen for their use of the \textsc{Maven} build tool, inclusion of developer-written \textsc{JUnit5} tests, 
no documentation of flakiness of tests,
minimal dependence on multi-threading and external mocking libraries, and finally, compatibility with JDK 8/11 and the XStream library. 
Each column of Table \ref{table:subjects} provides:
(1) subject project;~
(2) lines of code;~ (3) number of tests;~(4) number of analyzed mutants;~(5) number of analyzed test runs;~ (6) average number of covering tests for each analyzed mutant;~ (7) the mutation score for covered mutants; and (8) time taken by our experiment for the subject.
Across the 10 subjects, we analyzed fine-grained memory data from over 1.2 million mutated test runs for 46,958 mutants.
We found no test flakiness in the subjects' test suites.

\mysubsection{Experiment Running Time}
Our experiment ran on a 3.2Ghz Apple M1 ARM processor with 16GB RAM, and required 18 days (approx.) to complete, cumulatively for all 10 subject programs.
We designed the experiment to answer the research questions that we pose in this work.
Answering such research questions requires comprehensive runtime data, which we collect through the experiment's multiple stages: (a) running a program's 10 non-mutated runs; (b) running a program's mutated test run with instrumentation to collect runtime memory graphs; and (c) analyzing memory graphs for both surviving and killed mutants across all test runs.

Collectively, each such step incurs running-time costs.
In Table~\ref{table:subjects} we list these running times that vary with each program --- from as little as 1 hour 44 minutes (\spotify) to as much as 167 hours 36 minutes (\reader) --- often depending on the number of mutants and covering tests.

It is important to note that our experimental setup has sub-optimal performance and is not designed for use in practice. Our experiment was designed to test the scope and feasibility of our approach, and includes time-consuming, redundant steps to ensure the integrity of our experimental data.
For instance, we record memory data from test runs on both killed and surviving mutants, which are voluminous and incur substantial I/O and compression/decompression costs, to enable experimentation and exhaustive data analysis.
In practice, much of these expenses would not be needed.
Moreover, the analysis could be performed on individual or a smaller subset of mutants.
A tool developer could forego such costly redundancies when implementing our approach as a developer tool (\eg in an IDE or a CI/CD system), which assuredly would offering significant speed-ups.

\begin{table}[tb]
\footnotesize
\centering
\caption{Experimental Subject Programs}
\resizebox{\columnwidth}{!}{

\begin{tabular}{ |l|r|r|r|r|r|r|r|}

\hline
&         &        &         &        &   avg.   &           &   \\
Subject Project & KLoC    & \#T    & \#Mut   & \#Run  & \#CT     &    MS  & Time\\

\hline
commons-cli     & 6.2     & 381    & 723     & 36,084    &  49.9  & 92\% & 4h 56min\\ 
joda-money      & 9.2     & 1457   & 899     & 97,933    &  108.9 & 82\% & 12h 30min\\ 
cdk-data        & 10.6    & 4348   & 2,143   & 208,876   &  97.5  & 77\% & 49h 53min\\ 
jline-reader    & 13.3    & 180    & 2,920   & 130,700   &  44.8  & 66\% & 167h 36min\\ 
commons-validator  & 16.8    & 491    & 1,743   & 36,581    &  21.0  & 88\% & 8h 41min\\ 
commons-codec   & 24.1    & 1074   & 3,444   & 41,110    &  11.9  & 92\% & 14h 56min\\ 
spotify-web-api & 24.4    & 287    & 1,916   & 20,817    &  10.9  & 85\% & 1h 44min\\ 
commons-text    & 26.6    & 1241   & 4,258   & 75,461    &  17.7  & 85\% & 24h 07min\\ 
dyn4j           & 66.1    & 2308   & 9,736   & 286,366   &  29.4  & 78\% & 28h 01min\\ 
jfreechart      & 138.4   & 2306   & 19,176  & 337,159   &  17.6  & 59\% & 155h 16min\\

\hline
\end{tabular}
\label{table:subjects}

}
\vspace{-.2in}
\end{table}

 \section{results}
\label{sec:results}

\newcommand{\leading}{left\xspace}
\newcommand{\trailing}{right\xspace}

\begin{figure*}[t!]
  \centering
\subfloat[commons-cli]{
    \includegraphics[width=0.145\textwidth]{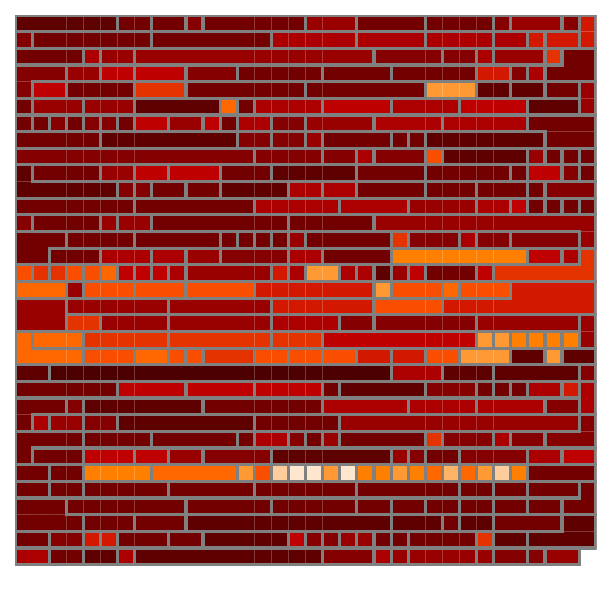}
    \includegraphics[width=0.145\textwidth]{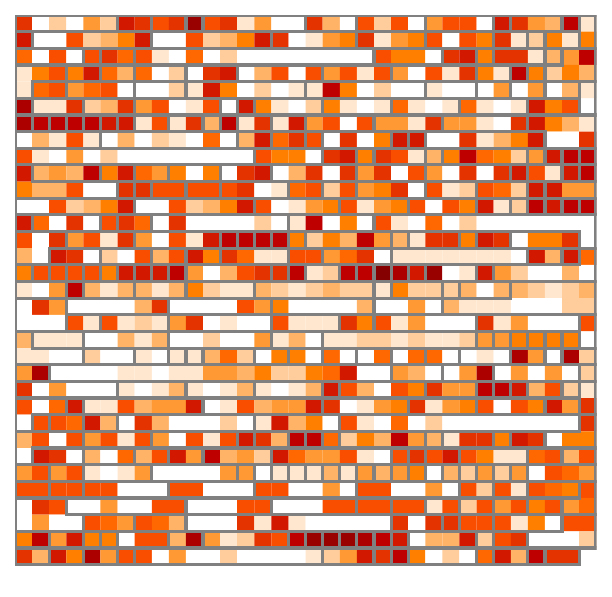}
    \label{fig:subfig12}
  }
  \subfloat[commons-codec]{
    \includegraphics[width=0.145\textwidth]{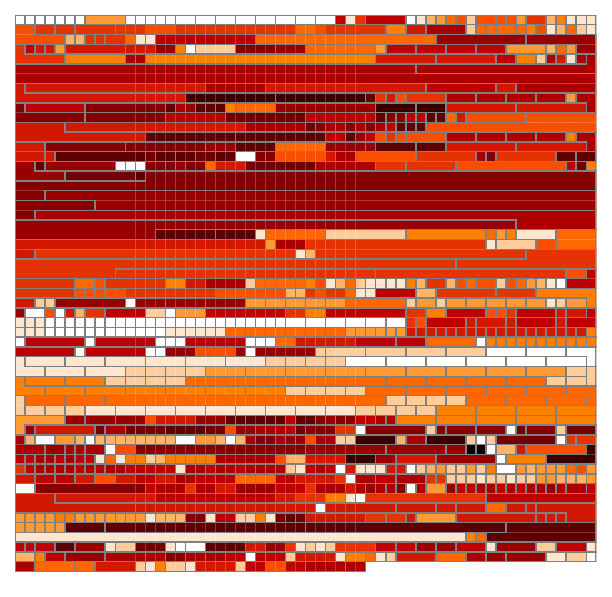}
    \includegraphics[width=0.145\textwidth]{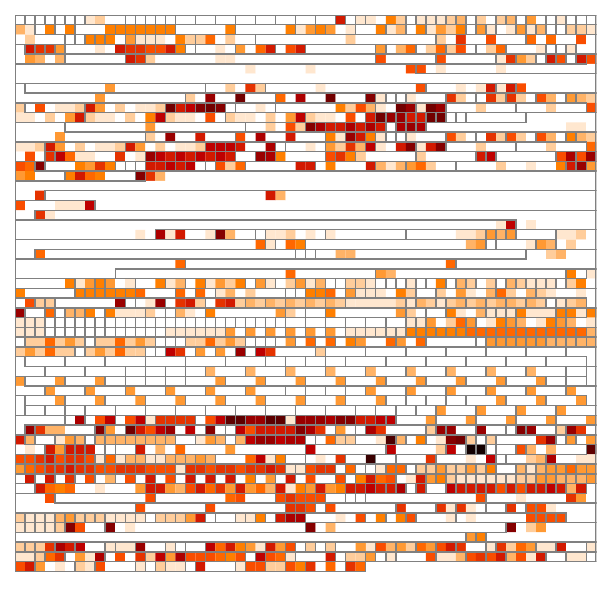}
    \label{fig:subfig13}
  }
\subfloat[commons-validator]{
    \includegraphics[width=0.145\textwidth]{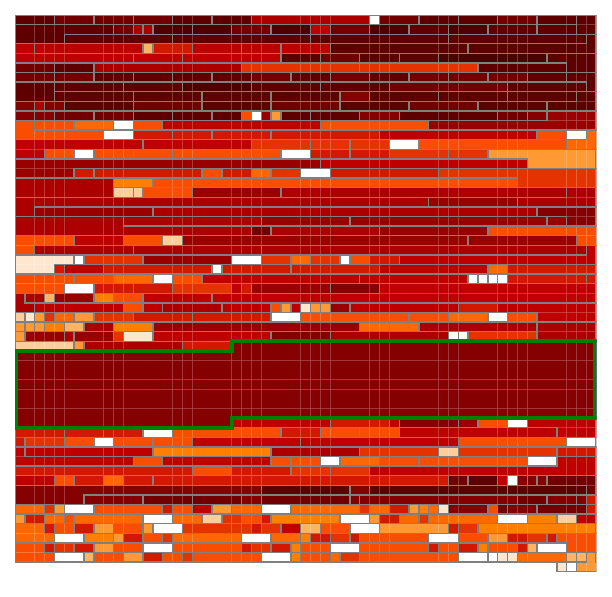}
    \includegraphics[width=0.145\textwidth]{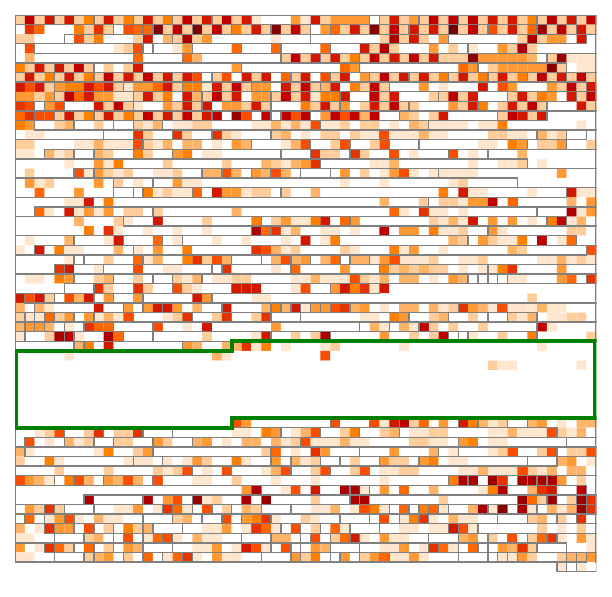}
    \label{fig:subfig15}
  }\\
\subfloat[jline-reader]{
    \includegraphics[width=0.145\textwidth]{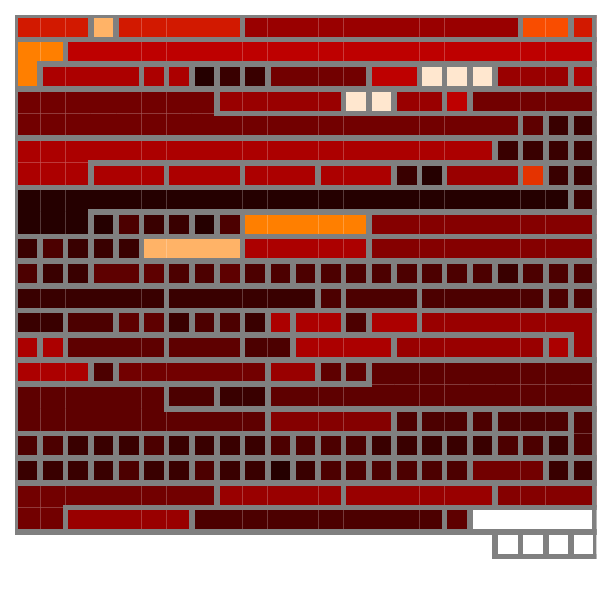}
    \includegraphics[width=0.145\textwidth]{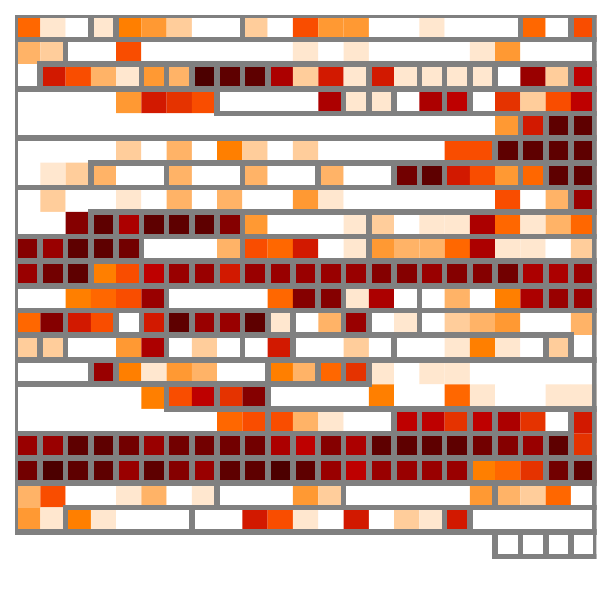}
    \label{fig:subfig-jlinereader}
  }
  \subfloat[joda-money]{
    \includegraphics[width=0.145\textwidth]{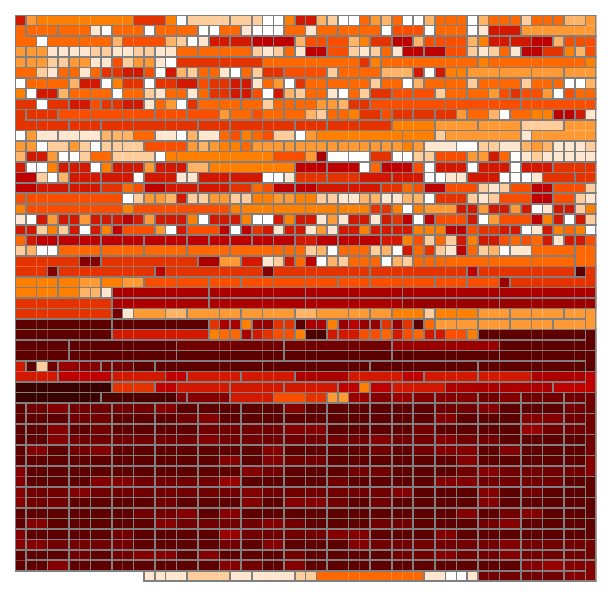}
    \includegraphics[width=0.145\textwidth]{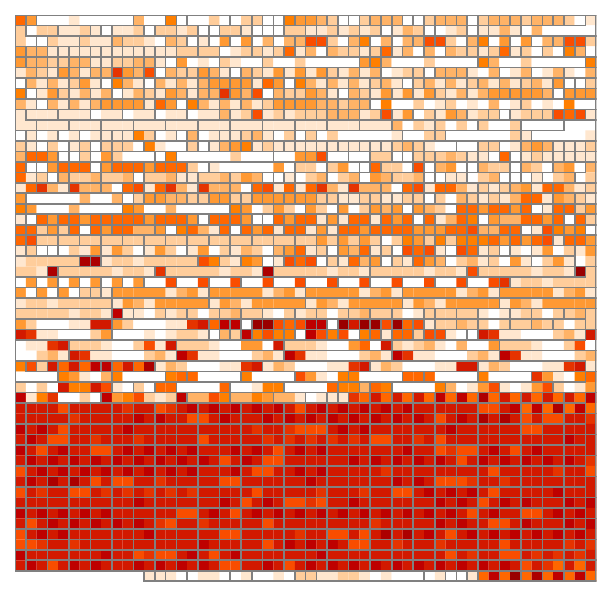}
    \label{fig:subfig23}
  }
  \subfloat[spotify-web-api]{
    \includegraphics[width=0.145\textwidth]{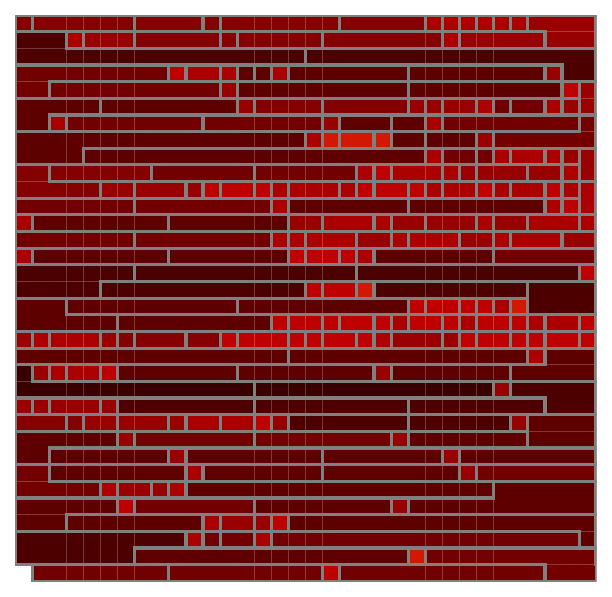}
    \includegraphics[width=0.145\textwidth]{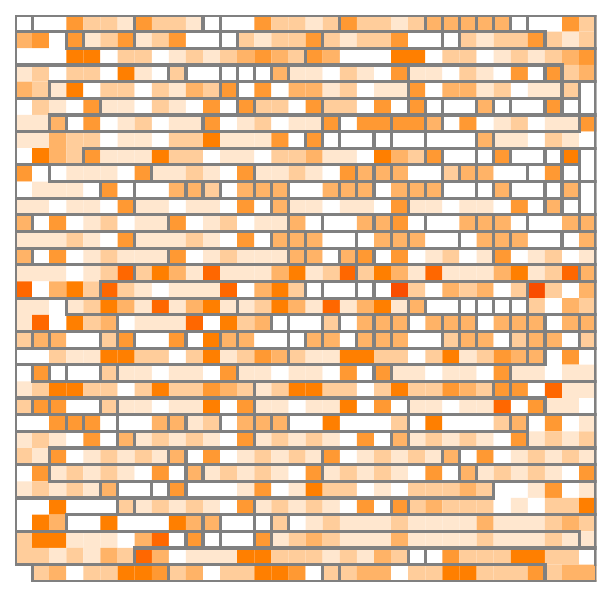}
    \label{fig:subfig24}
  }\\
\subfloat[Legend: mutant killing/crossfiring capability]{
    \includegraphics[width=0.50\textwidth]{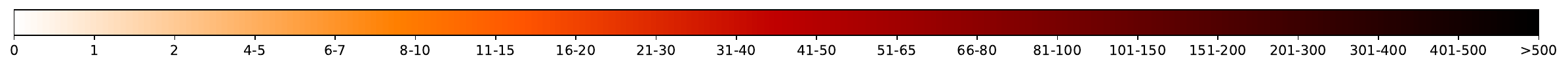}
    \label{fig:subfigbar}
  }
  \caption{Test and Assertion Crossfiring Capabilities}
  \label{fig:killed}
  \vspace{-.2in}
\end{figure*} 

\subsection*{RQ1: Test and Assertion Fault Detection for Killed Mutants}
In Figure \ref{fig:killed}, we employ a small-multiples approach~\cite{tufte2001visual} to illustrate varied capabilities of tests and assertions in killing mutants across 6 representative subject projects.
In each project, we present two sub-figures at the same scale: the \leading highlights the capability of test cases, while the \trailing delves into the granularity of individual assertions.
Test cases are visualized as regions outlined with gray borders.
Assertions are depicted as square pixels within the grey borders of their enclosing test case.
The size of a test's enclosed area (with the gray border) reflects the count of assertions within that test case.
The test cases and their assertions are sorted by their names and locations.

In the \trailing sub-figure for each project, each pixel's color indicates an assertion's mutant-killing capability, with darker, redder pixels denoting higher numbers of killed mutants (\ie greater crossfiring) by the corresponding assertion.
This color coding is also detailed in the legend in \Cref{fig:subfigbar}, where darker, redder shades indicate that an assertion or test has killed more mutants, while the lighter, more yellow shades suggesting ``fewer mutants killed,'' and white would indicate no mutant kills.
For the \leading sub-figure, the entire region of a test within a grey boundary is colored to reflect the overall strength of a test case, \ie the number of mutants each test case fails on, aggregated by test failures from all sources.
In other words, the color in these regions represents the crossfiring capability as measured at the test level.
When comparing mutant-killing capabilities, several key takeaways stand out: 
\begin{enumerate}
  \item The patterns formed by striped, rectangular regions observed in the \leading sub-figures for each subject reveal that the majority of developer-written test cases contain multiple assertions.
  \item Darker shades of red dominate the \leading sub-figures across all subjects. 
  These \leading sub-figures show mutant kill counts for individual test-cases.
  The dark red shades suggest the aggregated effects of all of a test's assertions' capabilities combined with its non-assertion failures' contributions.
\item In contrast, we see lighter shades of yellow and orange in the \trailing sub-figures. 
For tests' constituent assertions, some carry significantly more weight than others within a single test case or throughout the test suite. 
  In fact, many assertions detect zero faults (white) of the mutants used, whereas others in the same test case detect many.
  Overall, the sparse lighter-colored pixels at the assertion-level reveal a surprisingly smaller fraction and a more dispersed capability of individual assertions in their contributions to mutant kills.\end{enumerate}

Notably, \textit{testIPv6} in \validator (marked with green boundary in  \Cref{fig:subfig15}) stands out with 472 assertions and spans more than 500 lines in test code.\footnote{This may indicate a test smell instance because each individual assertion can be isolated as one single unit test case and parameterized.} 
This test case demonstrates a strong mutant-killing capability as a whole (deep red test region in the \leading subfigure), but only 28 assertions contribute to the mutation score.
Interestingly, the patterned bottom in \Cref{fig:subfig23} for \money shows the use of unit tests that share similar groups of assertions, which could be parameterized.

\begin{leftbar}
\noindent\textbf{Answering RQ1:} While a single test case may demonstrate a strong mutant-killing capability, the constituent assertions unevenly contribute to that capability, which reveals that some assertions are more capable of detecting more faults.

\end{leftbar}

\subsection*{RQ2: Killable Surviving Mutants}
We present the results for both \RQb and \RQc in Table~\ref{table:opportunities}, where 
we present summary statistics for assertion candidates for killing surviving mutants, as identified through our analysis. 
We group the columns of Table~\ref{table:opportunities} into three categories:
\begin{enumerate}
  \item Magnitude of killable surviving mutants: a count of surviving mutants (\#Surviving) alongside those that are killable (\#Killable), as identified by our analysis.
  \item Ways of killing each killable mutant: average number of ways to kill each killable mutant through checks by different (a) assertions ($\overline{\#{assert}}$), (b) test variables ($\overline{\#var}$), and (c) test cases ($\overline{\#test}$).
  \item Assertion Candidate Characteristics: a count of number of assertions to kill all killable, surviving mutants (\#Assert), along with the number of tests (\#Test) and test variables (\#Var) used in devising the assertions.
\end{enumerate}

\setlength{\tabcolsep}{4pt}
\setlength{\extrarowheight}{3pt}

\begin{table}[t]
\centering
\caption{Surviving Mutant-Killing Opportunities}
\resizebox{\linewidth}{!}{\begin{tabular}{ |l|r|r|r|r|r|r|r|}

\hline
& \multicolumn{1}{c|}{Magnitude of Killable} &  \multicolumn{3}{c|}{Ways of killing each} & \multicolumn{3}{c|}{Assertion Candidate}\\
& \multicolumn{1}{c|}{Surviving Mutants} &  \multicolumn{3}{c|}{killable mutant (avg)} & \multicolumn{3}{c|}{Characteristics}\\
\hline

Subject Project   &\#Killable$/$\#Surviving &  $\overline{\#{assert}}$ & $\overline{\#var}$ & $\overline{\#test}$ & \#Assert & \#Var & \#Test\\
\hline

commons-cli       &    36/60 (60\%)&     16& 15   &  13 &   312 &   298&   205\\
commons-valida. &   40/210 (19\%)&    224&   89   &  37 &  5405 &   531&   101\\
spotify-web-api   &   243/288 (84\%)&   99&  25   &  10 & 17414 &   563&   184\\
cdk-data          &   146/519 (28\%)&   196& 119  &  20 & 27341 & 10126&  1518\\
commons-text      &   298/644 (46\%)&    8&  7    &  4  &   941 &   848&   357\\
dyn4j             &  736/2188 (34\%)&  138&  19   &  6  & 25997 &  3039&  1000\\
commons-codec     &   65/284 (23\%)&    59&  18   &  4  &  3082 &   659&   175\\
joda-money        &   62/165 (38\%)&     6&  6    &  6  &   138 &   138&   137\\
jline-reader      &  414/1019 (41\%)&  509&  337  &  76 & 33871 &   946&   151\\
jfreechart        &  3464/8107 (43\%)&  78&  20   &  9  & 33488 &  2263&   844\\

\hline
\end{tabular}
}
\label{table:opportunities}
\vspace{-.1in}
\end{table}

\setlength{\tabcolsep}{6pt}
\setlength{\extrarowheight}{0pt}

The column labeled ``Magnitude of Killable Surviving Mutants'' in Table~\ref{table:opportunities} shows that our analysis is able to identify 19\%--84\% of surviving mutants as killable, depending on the subject program.
In other words, the existing tests already provide the necessary test data to execute, infect, and cause infection propagation back to at least one test; however, the tests do not have sufficient assertions to reveal it.
For instance, our memory-state analysis shows that 60\% of \cli's 60 surviving mutants are killable with assertion amplification to existing tests.
That number is as high as 84\% in the case of \spotify, where 243 out of 288 surviving mutants are killable.

\begin{leftbar}
\noindent\textbf{Answering RQ2:} For our subject programs, a fine-grained memory-state analysis reveals that 19\%--84\% of surviving mutants exhibit propagated infections not revealed by test cases, which can be killed through assertion augmentation of existing tests.
\end{leftbar}

\subsection*{RQ3: Opportunities in Tests for Assertion Amplification}
Given that many surviving mutants are indeed killable through assertion augmentation of existing tests, we next investigate the number of ways available (tests, test variables, assertions) to kill such mutants.

Again, consider the data in Table~\ref{table:opportunities}.
For each mutant, our technique is able to identify the total number of tests, test variables, and specific candidate assertions that can be of use in killing the surviving mutants.
Consider, \spotify's 243 killable surviving mutants.
We find that, on average, to kill a surviving \spotify mutant, our analysis can identify 99 assertion candidates that can be written using 25 different test variables (\eg local variables, method return values), which spread across 10 existing tests in \spotify's test suite.
Further, for all 243 killable mutants, our analysis detected $17,414$ killing assertion candidates, using 563 different test variables in 184 different existing test methods in \spotify's test suite. 

Across all subject projects, we identify comprehensive candidate solutions (assertions, tests, variables) to kill surviving, killable mutants through our analysis.
The extensive mutant-killing opportunities arise from the fact that a single surviving mutant may produce a varied magnitude (avg. 118 locations) of propagation on multiple test cases (avg. 14 tests), which can be accessed through multiple variables (avg. 46 variables).

Furthermore, we additionally assessed the average depth for an assertion to access a specific field (whose value can be asserted or evaluated) from a variable in the test code. We found that the average depth ranges from $1.5$ to $12.1$ across all subject projects.
For example, access paths that are four nodes deep suggest accessing heap locations through complex series of memory dereferences (\eg \texttt{var1.f1.f1.f3.f4}).
Long access paths can make assertions hard to read and brittle, leading to test anti-patterns. 
Therefore, it is important to filter out assertion candidates with excessively long access paths.

\begin{leftbar}
\noindent\textbf{Answering RQ3:} Our \analysis uncovers extensive assertion-amplification opportunities for each killable surviving mutant: on average, each can be detected through 118 infected locations across 14 tests in test code, while only one location in one test would be necessary to kill it.
  As such, developers would have numerous options to kill each mutant.
\end{leftbar}

\subsection*{RQ4: Optimizing Crossfire Strategies}

\begin{table*}[t]
\centering
\caption{Surviving Mutant Killing Strategies}

\resizebox{\linewidth}{!}{

\begin{tabular}{ |l|r|r|r|r|r||r|r|r|r||r|r|r|r|}

\hline
Subject Project& \#Kill & Dep & \multicolumn{3}{c|}{Strategy 1 (Assertion-Greedy)} & \multicolumn{3}{c|}{Strategy 2 (Variable-Greedy)} & \multicolumn{3}{c|}{Strategy 3 (Test-Greedy)}\\
\hline

 & &  & \textbf{\#Assert (factor)}  & \#Var & \#Test & \#Assert & \textbf{\#Var (factor)}  & \#Test &\#Assert & \#Var & \textbf{\#Test (factor)}\\
commons-cli       &   36& 1.4&\textbf{  27.0 (1.3)}&   26.9&   23.1&  27.0&\textbf{   26.0 (1.4)}&   23.0&  27.0&   26.9&\textbf{   14.0 (2.6)}\\
commons-valid. &   40& 3.4&\textbf{  24.0 (1.7)}&   23.4&   17.3&  24.0&\textbf{   16.0 (2.5)}&   11.8&  24.0&   19.8&\textbf{   10.0 (4.0)}\\
spotify-web-api   &  243& 2.5&\textbf{ 187.0 (1.3)}&   66.5&   48.6& 187.0&\textbf{   31.0 (7.8)}&   27.1& 188.0&   36.9&\textbf{   23.0 (10.6)}\\
cdk-data          &  146& 2.2&\textbf{ 109.0 (1.3)}&  105.8&   90.0& 109.2&\textbf{   92.0 (1.6)}&   81.1& 112.9&  104.2&\textbf{   66.0 (2.2)}\\
commons-text      &  298& 1.2&\textbf{ 235.0 (1.3)}&  234.9&  171.4& 235.3&\textbf{  233.0 (1.3)}&  170.4& 239.4&  238.8&\textbf{  161.0 (1.9)}\\
dyn4j             &  736& 1.7&\textbf{ 348.0 (2.1)}&  341.0&  248.8& 354.2&\textbf{  309.0 (2.4)}&  236.1& 357.3&  337.9&\textbf{  219.0 (3.4)}\\
commons-codec     &   65& 1.4&\textbf{  41.0 (1.6)}&   40.4&   39.4&  41.0&\textbf{   40.0 (1.6)}&   39.4&  42.0&   41.4&\textbf{   38.0 (1.7)}\\
joda-money        &   62& 1.6&\textbf{  40.0 (1.6)}&   40.0&   40.0&  40.0&\textbf{   40.0 (1.6)}&   40.0&  40.0&   40.0&\textbf{   40.0 (1.6)}\\
jline-reader      &  414& 3.4&\textbf{  81.0 (5.1)}&   69.8&   51.1&  88.0&\textbf{   49.0 (8.4)}&   38.8&  85.5&   55.5&\textbf{   31.0 (13.4)}\\
jfreechart        & 3464& 2.7&\textbf{ 555.9 (6.2)}&  473.2&  332.2& 559.2&\textbf{  384.6 (9.0)}&  309.1& 567.4&  446.4&\textbf{  298.0 (11.6)}\\
\hline
\end{tabular}
}
\label{table:crossfire}
\vspace{-.1in}
\end{table*}

In \Cref{table:crossfire}, we present the performance of three mutant killing-strategies introduced in \Cref{sec:approach}.
In the first three columns, we present project's name, the number of killable surviving mutants (\#Kill) as a baseline, and the average depth (Dep) of an assertion's access path, when accessing the runtime state, from a variable in test code.
In \Cref{fig:graph} we show examples of access paths; where for instance, \texttt{var2.f4.f3} is an access path, usable in an assertion, and accessed from the test code's local variable \texttt{var2} with a depth of 3.

For each strategy (assertion-, variable-, test-greedy), we provide three crossfire performance metrics, including the number of assertions (\#Assert), variables (\#Var), and tests (\#Test) to kill the baseline number of surviving mutants from 20 separate runs. 
Each strategy's best performance metrics are bolded with a crossfire factor included in the parentheses, which is the average mutant-killing capability of the test-case element.

Take \spotify as an example: iteratively searching for the most capable assertion (Strategy 1: Assertion-Greedy) yields a solution with an average of 187 assertions that can be devised using over 66 test variables across more than 48 test cases.
The 187 assertions would kill all 243 killable, surviving mutants where each assertion on average kills $1.3$ (\ie the crossfire factor) surviving mutants.

Similarly, iteratively searching for the most capable test-variable check (Strategy 2: Variable-Greedy) yields a solution that, on average, yields the same number of assertions (187) as in Strategy 1, but only requires a focus on 31 variables across 27 test cases.
Further, devising assertions using any one of those 31 test variables would, on average, kill $7.8$ surviving mutants (crossfire factor).

Likewise, iteratively searching for the most capable test case (Strategy 3: Test-Greedy) produces a solution that only requires augmenting 23 test cases, where each assertion-augmented test on average kills $10.6$ surviving mutants.
Moreover, the average depth of assertions' access paths is $2.5$ for killable mutants as a result of filtering out all assertions with non-shortest depths --- much shorter than the access-path depth of $5.2$, which is the average depth when not filtering out non-shortest-depth access paths.

Across all subjects, we observe significant mutant-crossfire effects enabled by our optimizing technique.
Each assertion can crossfire as many as $6.2$ surviving mutants in \jfreechart with the assertion-greedy strategy, and each assertion-augmented test can crossfire as many as $13.4$ surviving mutants in \reader with the test-greedy strategy.

Such leveraged crossfire effects can enlarge each incremental assertion amplification effort and scope the overall engineering and validation efforts within a few assertions, variables, or tests.
For example, our analysis found assertion amplification opportunities within $4,572$ tests (from \RQc) across all subjects, while the test-greedy strategy can scope the number down to $900$ tests (by aggregating the last column in \Cref{table:crossfire}).
As a result, only $1,684$ out of $147,989$ assertion candidates are selected.

Moreover, each candidate assertion to kill surviving mutants has the shortest depth of access path to assert the state, which on average ranges $1.2$--$3.4$, as compared to $1.5$--$12.1$ for candidate solutions before any of our three different optimizations.
Such a reduction may mitigate rendering hard-to-read and brittle assertions that check attributes deeply accessible from a variable.

When comparing the three optimizing strategies, we find that the test-greedy strategy (Strategy 3) is able to substantially reduce the number of tests to be improved.
Remarkably, for \spotify, the strategy almost halves the number of target tests (from 48.6 to 23.0) and test variables (from 66.5 to 36.9) that may require consideration for assertion amplification, with a negligible rise in the number of assertions (from 187.0 to 188.0).
Such significance reduction in the target tests can also be observed in multiple projects, such as \cli, \validator, \spotify, \cdk, and \reader, while not significantly increasing the number of assertions or variables to be checked for all subjects.

\begin{leftbar}
\noindent\textbf{Answering RQ4:} Our mutant-crossfire techniques with the test-greedy strategy can scope the amplification efforts down to $1,684$ from $147,989$ assertion candidates, while retaining the ability to kill all $5,504$ surviving, killable mutants across all ten subject programs.
\end{leftbar}

 \section{Discussion}

Our experimental results from the previous section offer several key insights into how we do test amplification, the characteristics of killable surviving mutants and cross-firing effects of tests, which we discuss next. 
We also offer a qualitative exposition of assertion candidates that we generate, using real examples from our experimental data.

\mysubsection{Key Takeaways from Experimental Data}
Our approach for assertion amplification looks to kill surviving mutants.
We target surviving mutants because we find that many such hard-to-kill mutants propagate program-state infections that go undetected by existing tests.
In our experimental subjects, we note that 19\%--84\% of surviving mutants can be killed with such additional mutant-killing assertions.

We also observe that multiple existing test cases can be amplified with additional assertions to kill surviving mutants.
Interestingly, a single surviving mutant can exhibit multiple state infections across various test cases: for each surviving mutant, on average, our memory-state analysis can detect 118 different infection locations across 14 existing tests.

We speculate the reasons behind such ``broad-spectrum'' killability of surviving mutants: software tests are intentional in their design and typically do not cover every possible fault in a program. 
They reflect a software tester's intent and priorities --- \eg the behaviors they are testing for, or the faults they are guarding against.
As such, the fact that multiple test cases (on average) produce detectable infection from surviving mutants may be an indication that these mutations are relevant to the test data that were chosen by the software’s developers, but the developers simply did not provide sufficient assertions to detect some relevant fault-revealing attributes. 
We further surmise that such surviving mutations likely mimic faults that are meaningful to the program's logic and semantics, as well as its human-written test data and harnesses.
Therefore, such surviving mutants may warrant additional assertions, to guard against faulty program behaviors.

Critically, we again observe this ``broad-spectrum'' killability when studying crossfiring properties of tests and their constituent parts: in many cases, a single test assertion, test variable, or test case can detect and kill multiple mutants.

We leverage this observation to scope the amplification efforts within only a few test cases that we might augment, which can bring down the number of assertion candidates from $147,989$ to $1,684$ among $900$ tests, while still being able to kill all $5,504$ surviving, killable mutants across the ten subject programs.
This provides a crossfire factor of $6.1$ (\ie $900$ tests crossfiring $5,504$ surviving mutants).

\begin{figure}[t!]
    \centering
    \includegraphics[width=\linewidth]{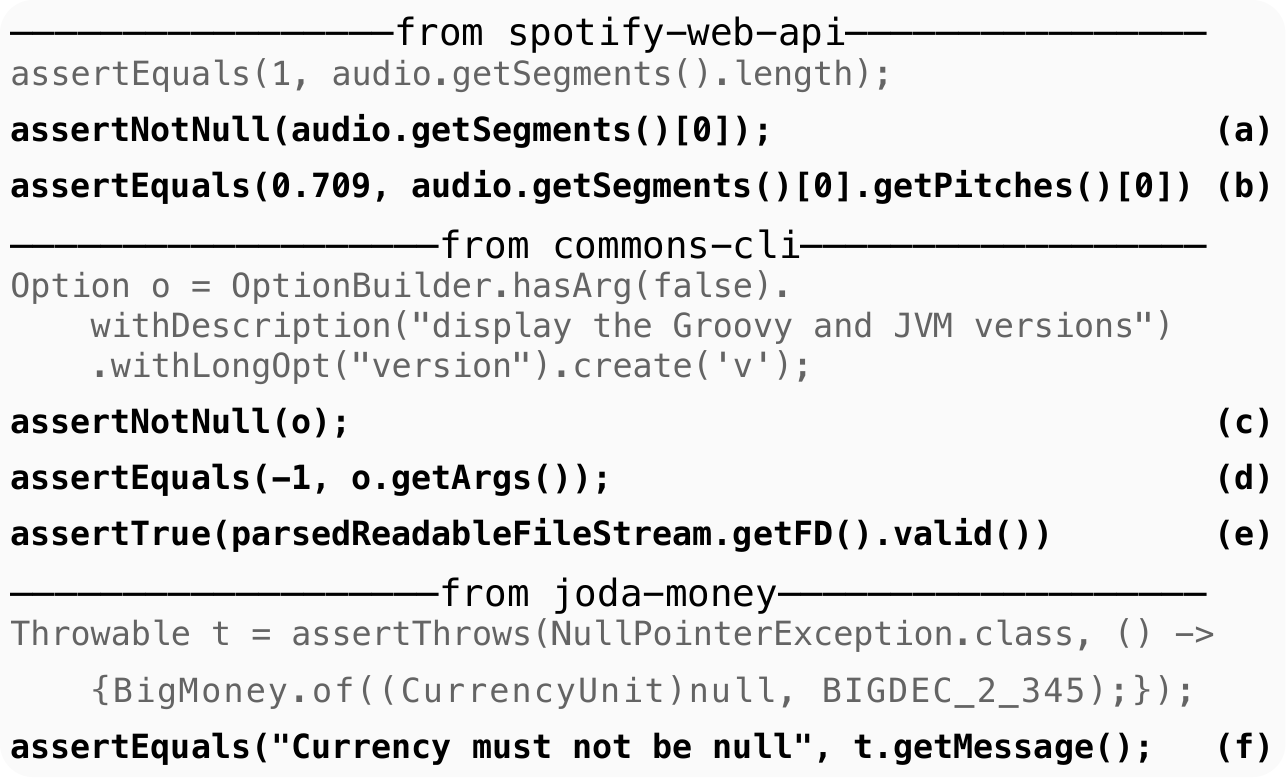}
    \caption{Assertion Examples}
    \label{fig:code}
    \vspace{-.2in}
\end{figure}

\mysubsection{Representative Examples of Assertion Candidates}
In Figure~\ref{fig:code}, we enumerate six representative examples of assertion candidates generated from our experiments on three subject programs: \spotify, \money, and \cli. 
The assertion candidates are highlighted in bold and presented alongside snippets of the original test code. 
Each assertion candidate is marked as (a) through (f).

These examples offer insights into the characteristics and suitability of assertion candidates produced by our approach. 
While not exhaustive, they serve as a qualitative exposition of our technique and results. 
To capture a diversity of assertion candidates, we include examples that (a) might be adopted by developers or (b) might introduce test anti-patterns. 
We also showcase how our approach safeguards against assertion candidates that might introduce test anti-patterns.

\mysubsubsection{Examples that might be adopted by developers}
Consider the Assertion Example (a) from \spotify and (c) and (d) from \cli.
We speculate that such assertions are likely to be adopted by developers, due to their simplicity.

These assertions check a variable directly (\eg (c) performs a null check on test variable \texttt{o}) or have shallow access paths to attributes, such as fields defined in project production code (\eg \texttt{o.getArgs()} in (d) or \texttt{audio.getSegments()[0]} in (a)).

For our experimental subjects, we find that on average, 42.1\% of the assertion candidates (using the test-greedy strategy) simply check a test variable directly, including checking primitive values, string values, variable type, or their nullability.
We also observe that 41\% of assertion candidates access first-party attributes from production code, via fields or accessor methods (\eg \texttt{getArgs()}).

Further, assertion examples in (a), (c) and (d) test variables and states in first-party production code, without requiring access to third-party or system code.
On average, 85\% of all assertion candidates from our experiments access variables and program states from first-party production code.

\mysubsubsection{Examples that might introduce test anti-patterns}
Consider the example assertions (e) in \cli and (f) in \money.
These assertions might introduce test anti-patterns for different reasons, and as such might not be adopted by developers.
Assertion Example (e) checks an attribute from an external library via a deep access path to examine the validity of a file descriptor (\texttt{.getFD().valid()}). 
Assertion Example (f) checks the value of a method return (\texttt{t.getMessage()}) from a system-level class (\texttt{Throwable}). 
We hypothesize that developers might avoid writing tests for external library code, especially with deep and specific access paths, leading to a higher likelihood of inducing test anti-patterns due to reliance on external libraries and specificity.

Notably, as a side-effect our crossfiring-based filtering strategy (see Section~\ref{sec:step2}, and \RQd experiment results in Section~\ref{sec:results}) reduces the average depth of access-paths for assertion candidates from 6.3 to 2.2.
As such, when optimizing for cross-firing effects, our approach mitigates the likelihood of selecting assertion candidates with deep, complex access paths that might lead to test anti-patterns.

For instance, between Assertion Examples (a) and (b) in \spotify that both detect the same mutant, when optimizing for the number of tests (\ie test-greedy strategy shown in results for \RQd), our approach would opt for assertion candidate (a) --- the simpler assertion candidate via shorter access paths. 

\mysubsection{Future Work}
We acknowledge that our insights into the selected assertion candidate examples depend on the authors' categorization methods and do not capture actual developer sentiment or feedback. 
The actual suitability of each assertion candidate is subject to the specific test case context, testing, and project requirements.
Moreover, we recognize that our techniques provide multiple options for each surviving mutant, while only selecting one based on our strategy.
Offering developers multiple choices allows them to select the most suitable assertion candidates or make decisions to isolate a new test, thus reducing the occurrence of anti-pattern assertions. 
In the future, we will conduct user studies to evaluate the suitability of these assertion candidates and design tools that investigate human aspects in mutation-testing research.

 \section{Threats to Validity}

The external threats stem from the generalizability of our findings:
our observations on test/assertion-level mutant detection, characteristics of unrevealed propagation, and the performance of mutant-crossfire strategies may not apply to other mutation operators, subject projects, or programming languages. 
However, we use a popular mutation-testing framework, PIT \cite{PIT}, with all its default group of mutation operators, and select subject projects that vary in production-code size, test-suite size, and complexity.
Further, across all 10 of our subjects, we saw consistent observations: every subject has many surviving mutants that are killable through mere assertion augmentation of existing tests, and that crossfiring was a source of substantial savings, in terms of the number of assertions needed to kill all killable, surviving mutants.

Internal validity is challenged by the presence of duplicate mutants, which PIT addresses through using a restrictive set of mutation operators that prevent operator subsumption, ensuring a one-to-one mapping between mutants and target syntax, and filtering out byte-code identical equivalent mutants \cite{Coles_2023}. 
Execution nondeterminism is another concern. 
We mitigate it by our use of 10 repetitive no-mutation test runs, followed by analysis of test consistency through object-graph walks, and isolation of each mutant's runs in separate JVM instances with static-field cleaners.

Finally, the construct validity threats come from the limitations of our evaluation metrics and experimental setup.
Conducting human-involved studies or submitting pull requests to project maintainers would gain further insights into the actual suitability of our suggested assertion candidates.
A challenge to such a study of pull-request acceptance rates is motivating the pull-request to project maintainers who have no experience nor knowledge of mutation testing---\ie we must answer the question, ``why are we submitting a test that catches a bug that does not exist?'' for developers who may not be aware of such approaches.
In this study, we demonstrate the potential benefits of using different crossfire strategies, present our assertion candidates' characteristics, and discuss assertion candidates' suitability.
In the future, we will conduct human studies to further explore the suitability of our approach.

 \section{Related Work}

\mysubsection{Mutant Crossfiring in Mutation Testing}
Recent literature distinguishes between mutation \emph{analysis} and mutation \emph{testing} \cite{Kaufman2022PrioritizingMutants,PAPADAKIS2019275}: mutation analysis assesses the strength of the test suite, while mutation testing focuses on resolving each surviving mutant to strengthen the test suite. 

Smith and Williams \cite{SmithWilliam2007Crossfire,smith2009guiding} documented mutant ``crossfire,'' where a new test targeting one surviving mutant coincidentally kills others. Jia and Harman \cite{JIA2009Subsuming} noted that mutants can be ``collaterally'' killed by tests aimed at different mutants. This phenomenon, where a single test kills multiple mutants, appears in mutation-testing research on mutation redundancies (\eg \cite{ToKillAMutant2023Hang,Just2012Redundant,KAMINSKI20132002,Just2012Nonredundant,Papadakis2015TCE}), mutant-subsumption relationships (\eg \cite{JIA2009Subsuming,kintis2010evaluating,Papadakis2016subsuming}), and mutant-ranking techniques (\eg \cite{titcheu2020selecting,Kaufman2022PrioritizingMutants}).

We also observe and confirm the crossfire phenomenon in mutation testing from our investigations. Unlike prior studies that analyze mutants, we offer a model to examine the causes of this phenomenon at fine-grained memory-state levels. 
Furthermore, we use these insights to develop and prioritize crossfiring assertion candidates to target crossfiring mutants.
Moreover, we recognize that practitioners often work with individual surviving mutants and perform incremental augmentation in empirical mutation-testing research, both in academia (\eg \cite{smith2009guiding, Kaufman2022PrioritizingMutants, Rojas2017GamifyCroudsource, rojas2016code}) and in industry (\eg \cite{petrovic2018state, Petrovic2023FixThisMutant, beller2021would}). 
This focus informs our work, which 
differs from other test-generation works in many aspects.

\mysubsection{Test Generation and Amplification}
Past efforts in test generation include test-data generation (\eg \cite{Souza2014TestDataMutationTesting,Anand2013SurveyTestDataGeneration,chekam2021killing}), assertion generation in programs (\eg \cite{Ernst2001Invariant,Boshernitsan2006Agitator,Terragni2021Gassert,Boshernitsan2006Agitator}) or test code (\eg \cite{Xie2006Augment,Zamprogno2023Human,staats2012automated,Tiwari2024Mimic}), and whole test-suite generation (\eg \cite{Fraser2013WholeTestSuiteGeneration,Fraser2011EvoSuite,Fraser2012Generation}). 
Some of these efforts are mutation-based (\eg \cite{titcheu2020selecting,Fraser2012Generation,fraser2010mutation,staats2012automated,veraperez2019suggestionstestsuiteimprovements}). 
Additionally, Vera-Pérez proposed a mutation operator that suggests improvements into test suites with infection and propagation analysis \cite{veraperez2019suggestionstestsuiteimprovements}

Previous work has also explored assertions' roles in mutant killing \cite{ToKillAMutant2023Hang} and test suite effectiveness \cite{zhang2015assertions}, brittle assertions \cite{Huo2014BrittleAssertion}, and realistic tests \cite{Bozkurt2011InputReal,Fraser2011Realisitc}.

Many works exploit existing tests and are considered test amplification \cite{Danglot2019Amplification}. Danglot \etal~\cite{Danglot2019Amplification} conducted a snowballing literature study and categorized test amplification into four types: adding new tests, synthesizing tests for changes, modifying test execution, and modifying existing test code. Our work falls into the last category but differs from prior works in key ways.

Existing test generation and amplification work primarily aim to achieve a broader goal of improving test suites, while using coverage/mutation score as a proxy fitness-function.
For example, Fraser and Arcuri \cite{Fraser2011EvoSuite,Fraser2013WholeTestSuiteGeneration} developed \textsc{EvoSuite} to generate an entire test suite from source code; Baudry \etal~\cite{Bandry2015Dspot,danglot2019dspot} created \textsc{Dspot} to perform test amplification by exploring more input space and assertions to enhance human-written tests. 
In contrast, our approach targets individual surviving mutants in mutation testing, and prescribes assertion candidates to kill such unkilled mutants. 

Further, to target individual surviving mutants, our approach conducts a comprehensive memory-graph walk to identify specific differences in memory state, yielding precise assertion candidates (\eg asserting a specific element in an array rather than the entire array). Prior test generation/amplification techniques (\eg~\cite{danglot2019dspot,terragni2020evolutionaryGassert,Fraser2013WholeTestSuiteGeneration}) compare primitive or string-type values, or entire objects to capture program states for overall test-suite improvement. 
In contrast, we capture the full object state, detecting infections in any enclosed fields, arrays, or collections accessible from an object. 
Additionally, we mitigate brittle assertions by selecting those that check shallow attributes of a variable while rendering minimal and precise assertion candidates.

Finally, previous techniques apply separate heuristics to minimize tests, such as reducing the number of generated assertions \cite{Fraser2011EvoSuite} and prioritizing the most effective tests \cite{danglot2019dspot}. 
Our amplification technique targets individual mutant killing and as such expects incremental code changes by practitioners. 
Therefore, instead of selecting a single minimization approach, we develop and compare different strategies that scope the incremental amplification efforts to a few assertions, variables, or test cases that ultimately achieve the same mutation score.

\section{Conclusion}
\label{sec:conclusion}

In this work, we developed a technique to identify assertion-amplification opportunities for surviving mutants via memory-state analysis. We found that up to 84\% of surviving mutants can be killed by reusing existing tests and augmenting their assertions. Each surviving mutant offers multiple locations in the test code for potential assertion amplification. Additionally, we used the phenomenon of mutant crossfiring, as encountered by practitioners, to construct a theoretical model and offer empirical insights for crossfiring at various granularities. 
Building on these insights, we devised a technique that optimizes the amplification process, with which we find that we can kill all such killable surviving mutants detected from our analysis with fewer tests and fewer assertions, providing a crossfire factor of 6.1x.
In future work, we aim to conduct human-centered studies and validate our approach with mutation-testing practitioners.

 \section*{Acknowledgments}
The second author's opinions expressed in this publication are solely his and do not purport to reflect the opinions or views of his employer, Microsoft.

\bibliographystyle{IEEEtran}
\IEEEtriggeratref{43}
\bibliography{IEEEFull}

\end{document}